\documentstyle[12pt]{article}
\begin{document}
\title{ General Relativistic Relation Between Density Contrast and Peculiar
Velocity}
\author{Reza Mansouri\footnote{E-mail:mansouri@theory.ipm.ac.ir} \\
Sohrab Rahvar\footnote{E-mail:rahvar@theory.ipm.ac.ir}\\
Department of Physics, Sharif University of Technology,\\
 P.O.Box 11365--9161, Tehran, Iran\\
Institute for Studies in Theoretical Physics and Mathematics,\\
P.O.Box 19395--5531, Tehran, Iran} \maketitle

\begin{abstract}
Concepts like peculiar velocity, gravitational force, and power
spectrum and their interrelationships are of utmost importance in
the theories of structure formation. The observational
implementation of these concepts is usually based on the
Newtonian hydrodynamic equations, but used up to scales where
general relativistic effects come in. Using a perturbation of FRW
metric in harmonic gauge, we show that the relativistic effects
reduce to light cone effects including the expansion of the universe.\\
Within the Newtonian gravitation, the linear perturbation theory of large
scale structure formation predicts the peculiar velocity field to be
directly proportional to gravitational force due to the matter distribution.
The corresponding relation between peculiar velocity field and density
contrast has been given by Peebles. Using the general relativistic
perturbation we have developed, this familiar relation is modified by
doing the calculation on the light cone in contrast to the usual
procedure of taking a space-like slice defined at a definite time. The
velocity and density--spectrum are compared to the familiar Newtonian
expressions. In particular, the relativistic $\beta$--value obtained is
reduced and leads to an increased bias factor or a decreased expected amount
of the dark matter in a cluster.
\end{abstract}
\section{Introduction}
Although cosmology has been one of the first areas of application
of general relativity, in physical cosmology we encounter rarely
general relativistic considerations. Relativistic effects are
commonly restricted to either areas of very high density, like
black holes or neutron stars, or global cosmological effects,
like global dynamics of FRW models. In theoretical
implementations of observational cosmology, such as theoretical
interpretation of the Two-degree Field(2dF)-- and the Sloan
Digital Sky Survey (SDSS)--projects, the Stromlo-APM red-shift
survey \cite{tad96} and the Las Campanas red-shift survey
\cite{she96} one usually rely on basic Newtonian dynamics
\cite{pee80}, \cite{ber89}, \cite{pad93} and \cite{man00}. The
correct theoretical interpretation of these surveys needs a
through understanding of structure formation processes in the
universe, mainly objects of scales greater than galaxies and at
cosmological distances. It is obvious that at such scales one can
not use the full formalism of general relativity to study the
dynamics of structures and one has to rely on some approximation
methods. In relativistic jargon, the Newtonian limit means
gravity in the vicinity of objects having 'small' mass, or
expressing it more exactly, where the ratio of the Schwarzschild
radius to object radius is a very small number \cite{mis78}. But
one never discuss the relativistic effects on cosmological
parameters in the intermediate scales where the relativistic
effects due to the mass concentration is negligible. This is
exactly the case in the theory of structure formation. For small
density contrasts it is used to apply the Newtonian dynamics to
the linear theory of structure formation. But even if one tries
to apply the linear theory of structure formation to evaluate the
observational data, general relativistic effects due to the large
extension of the objects or their distances come in. This is exactly
the point which has been neglected in
theoretical studies up to now and we are going to consider it. \\

The problem is to formulate an approximation method within
general relativity for the cases where we have small density
perturbation in a FRW model for intermediate regions smaller than
the Hubble radius but large enough to be forced to consider light
cone effects. Specifically, we will calculate the relation
between the density contrast and the peculiar velocity in
clusters within the general relativity and compare it to the well
known Newtonian relations. The formalism developed here shed some
light on the methodology of the calculation of different so-called
light cone effects which has been published \cite{nis99}, \cite{mat97},
\cite{nak98}, \cite{de98} , \cite{mos98}.
The relativistic calculation helps us to understand more
systematically the complex role of different cosmological
parameters which come in once. In Section $2$ we review the basic
hydrodynamical equations based on the Newtonian gravity. In
Section $3$ a perturbation formalism for FRW universes in harmonic
gauge is given. Using this approximation formalism, we obtain a
relativistic relation between the density contrast and the
peculiar velocity. In Section $4$ we obtain relativistic derivation from
density contrast. Section $5$ is devoted to the
power spectrum. The general relativistic $\beta-$value is then
calculated in Section $6$. And finally in Section $7$ we summarize
the main points of this study.\\
Throughout the paper we choose the
signature $(-,+,+,+)$, and assume $c = 1$, except where indicated.

\section{Peculiar Velocity Field in the Newtonian Hydrodynamics}
Basic hydrodynamical equations used to study the peculiar velocity
of structures in the linear regime are the continuity and the
Poisson equation:
\begin {eqnarray}
\label{1}
\dot\delta+\nabla\cdot{\bf v} = 0,\\
\label{2} \nabla\cdot g = 4\pi G\rho\delta,
\end{eqnarray}
where $\delta=\frac{\rho-\rho_0}{\rho_0}$ is the density contrast in the
homogeneous and isotropic universe with the background density $\rho_0$. The
peculiar velocity is defined by ${\bf v(r) = {\bf u(r)} - H_0{\bf r}}$, where
${\bf u}$ is the velocity of cosmic fluid, $H_0$ is the Hubble constant, and
$r$ is the physical distance.
Within the Newtonian gravitational theory, the following relation between
the density contrast and the peculiar velocity is then derived by Peebles
\cite{pee80}:
\begin{equation}
\label{vq}
 \vec{v}(\vec{r})=\frac{H_0f(\Omega )}{4\pi b}\int
d^3r^{^{\prime }}\delta (r^{^{\prime
}})\frac{\overrightarrow{r^{\prime }}-\overrightarrow{r}}{ \left|
r^{^{\prime }}-r\right| ^3},
\end{equation}
where $b$ is the biasing factor defined by
$$\delta _{galaxy} = b\delta,$$
and $f(\Omega) = \frac{d\ln D}{d\ln a} \simeq {\Omega}^{0.6}$.
Eq.(\ref{vq}) can also be written as follows:
\begin {equation}
\label{peebles}
 \nabla \cdot \vec{v}=-\frac{H_0f(\Omega )}b\delta.
\end {equation}
Note that in general $f(\Omega)$ depends also on $\Lambda$. However it has
been shown
by Lahav \cite{lah91} that the effect of $\Lambda$ is negligible.

\section{Peculiar Velocity in the Relativistic Hydrodynamics}
Our aim is to study the peculiar velocity of a linear structures in
FRW space--time. The structure is assumed to be large enough to
take part in the cosmological expansion, yet smaller than the
present Hubble radius, with a density contrast small enough to
justify the use of the linear theory.

This is in contrast to the familiar textbook approach which uses Newtonian
gravity for the linear regime in the scales less than the Hubble radius. As
a result of our general relativistic calculations we will distinguish between
intermediate scales where relativistic effects in some reasonable
approximations have to be taken into account and small scales where the
Newtonian approximation is valid within the observational accuracy. Therefore,
we perturb linearly a FRW metric, or what is the same, we expand a general
metric linearly around a FRW space--time.  \\
Over the past decades, perturbation of FRW space--times have been
studied using different gauges like synchronous, Poisson and
restricted--gauge (for review see \cite{ber95}, \cite{muk92} and 
\cite{ber01}).
These gauges are, however, not suitable for our purposes. As we
know from the comparison of the weak limit of general relativity
to Newtonian gravity, the harmonic gauge is the most suitable one
playing the role of the Lorentz gauge in electromagnetism as
contrasted to the Coulomb gauge. Therefore, we try to formulate
the corresponding gauge in the
perturbed FRW space--time. \\
Consider a small perturbation of FRW metric in the form
\begin {eqnarray}
g_{\mu \nu}= g^{(0)}_{\mu \nu} + h_{\mu \nu}\nonumber\\
T_{\mu \nu}= T^{(0)}_{\mu \nu} + \delta T_{\mu \nu},
\end {eqnarray}
where $g^{(0)}_{\mu \nu}$ is the background FRW metric,
$T^{(0)}_{\mu \nu}$ the stress tensor of the cosmic fluid, $h_{\mu
\nu}<< g^{(0)}_{\mu \nu}$ the metric perturbation, and $\delta T_{\mu
\nu}<T^{(0)}_{\mu \nu}$ energy momentum tensor of the structures.
The structures we are considering are in a matter dominated
universe. Therefore, the only non-zero component of the energy
momentum tensor in the comoving reference frame is $
T^{(0)}_{00}=\rho_0 $. Taking into account the general form of the
energy--momentum tensor of a homogeneous perfect fluid:
\begin {equation}
T^{\mu \nu} = (p+\rho)u^{\mu}u^{\nu} + pg^{\mu \nu},
\end {equation}
where $u^{\mu}=\frac{dx^{\mu}}{d\tau}$ is the four-velocity of the perturbed
cosmic fluid. we may write its perturbed component as follows:
\begin{eqnarray}
\label{T0}
T_{00} &=& (\rho_0+\delta\rho)= \rho_0 (1 + \delta) \\
\label{T1}
T_{0i} &=& \rho_0u^{i}\\
\label{T2}
 T_{ij} &=& 0,
\end{eqnarray}
where we ignored higher order of perturbation. We introduce
Harmonic gauge in FRW metric as follows:
\begin{equation}
\label{har}
\frac{1}{2}h_{\mu}{}^{\mu}{}_{;\nu} =
h_{\mu\nu;}{}^{\mu}
\end{equation}
For the $\nu = 0$ and $\nu = i$ in the expression (\ref{har}), harmonic
gauge can be written in the FRW as follows:
\begin{eqnarray}
\label{nu0}
 \frac{1}{2}h^{0}{}_{0,0} + h^{i}{}_{0,i} -
\frac{1}{2}h^{i}{}_{i,0}+3\frac{\dot a}{a}h^{0}{}_{0}
-\frac{\dot a}{a}h^{i}{}_i = 0 \hspace{1cm} \nu=0 \\
3\frac{\dot a}{a}h^{0}{}_{i} - \frac{1}{2}h^{j}{}_{j,i} -
\frac{1}{2}h^{0}{}_{0,i}+ h^{j}{}_{i,j} - h^{0}_{i,0} = 0
\hspace{1cm} \nu=i
\end{eqnarray}
we obtain the (00)--component of the perturbed Einstein equation
$\delta G^{0}{}_{0} = 8\pi G\delta T^{0}{}_{0}$, after some
lengthly calculations, in the following form:
\begin {equation}
\label{g00}
 h_{0}{}^{i}_{,0i} + 3\frac{\dot a}{a}h_{0}{}^{i}_{,i}
- \frac{1}{2}h_{00,i}{}^{i} - 2\frac{\ddot a}{a}h^{i}{}_{i} -
2(\frac{\dot a}{a})^2h^{i}{}_{i} -\frac{1}{2}h^{i}{}_{i,00} -
3\frac{\dot a}{a}h^{i}{}_{i,0} - \frac{3}{2}\frac{\dot
a}{a}h_{00,0} - \frac{3}{2}(\frac{\dot a}{a})^2 h_{00} = 4\pi
G\rho \delta
\end {equation}
Substituting $\nu=0$ component of harmonic gauge expression (\ref{nu0})
into (\ref{g00}), Einstein equation obtain as follows :
\begin{equation}
\label{pos}
 -\frac{1}{2}h_{00,\mu}{}^{\mu} + 3\frac{\dot a}{a}h_{00,0}
 - \frac{1}{2}\frac{\dot a}{a}h^{i}{}_{i,0}
 + \frac{9}{2}(\frac{\dot a}{a})^2h_{00}+\frac{\ddot a}{a}(3h_{00}
 - h^{i}{}_{i}) = 4\pi G \delta\rho
\end{equation}
Note that for $a = constant$, i. e. the Minkowski space time, the
familiar Poison equation is recovered. Now, for scales smaller
than the Hubble radius, i.e. $\lambda < H^{-1} = \frac{a}{\dot
a}$, it is easily seen that all the terms in the left hand side of
(\ref{pos}) can be ignored relative to the first term. In
fact, we have to compare the Hubble time $t_H = \frac{a}{\dot a}$
with the characteristic time of changes in the cluster, $\tau$,
which is less that $\lambda$. It is then obvious that the second
,third  and fourth term which are of the order of $(t_H
\tau)^{-1}$ and fifth term is in the order of $t_H^{-2}$ are
ignorable relative to the first term which is of the order of
$\tau^{-2}$. Therefore, taking the definition of the
gravitational potential of the perturbing field, $\phi$, as
$h_{00}=-2\phi$, we obtain finally for the perturbed field
equation:
\begin {equation}
\Box \phi = 4 \pi G \rho\delta,
\end {equation}
where spatial derivatives of $\phi$ are in the terms of the physical
coordinates ${\vec {r}}$. We see that the dynamics of perturbing
potential obeys the same equation as in the case of relativistic
radiation, taking into account the difference between comoving
and physical--length. We may write therefore the potential in the
retarded form:
\begin {equation}
\phi(\vec {r}) = -\int \frac{G \rho(t - {\left |\vec {r'}-\vec{r}\right|})
\delta (t - {\left |\vec {r'}-\vec{r}\right|})}{\left |\vec{r'}-
\vec{r}\right|} d^3r'.
\end {equation}
Now, the gravitational acceleration $\vec g=-\vec\nabla\phi$ due to the above
gravitational potential is given by:
\begin {eqnarray}
\label{gg}
g(\vec{r})&=&-\int\frac{G(\vec{r'}-\vec{r})}{\left|\vec{r'}-\vec{r}\right|^3}
\rho(t-{\left|\vec{r'}-\vec{r}\right|})\delta(t-{\left|\vec{r'}
-\vec{r}\right|})d^3r'\nonumber\\
&-&\int\frac{G(\vec{r'}-\vec{r})}{c\left|\vec{r'}-\vec{r}\right|^2}
[ \dot\delta(t-{\left|\vec{r'}-\vec{r}\right|})
\rho (t-{\left|\vec{r'}-\vec{r}\right|})\nonumber\\
&+& \delta (t-{\left|\vec{r'}-\vec{r}\right|})
\dot\rho(t-{\left|\vec{r'}-\vec{r}\right|}]d^3r'.
\end {eqnarray}
Note that the background density in the FRW universe is
$\rho\propto a^{-3}$, and in the linear regime for the structure
formation, we have $\delta \propto a$. It is then easily seen that
the ratio of second term in (\ref{gg}) with respect to the
first one is of the order of $\frac{ \lambda}{d_H}$. Therefore,
the second term may be ignored and we obtain finally
\begin{equation}
\label{retard}
 g(\vec {r}) =
-\int\frac{G(\vec{r'}-\vec{r})}{\left|\vec{r'}-\vec{r}\right|^3}
\rho\delta(t-{\left|\vec{r'}-\vec{r}\right|})d^3r'.
\end {equation}
One may therefore interpret the gravitational effect of such linear
perturbations at intermediate scales as the Newtonian one with two
modifications:\\
i. All coordinate lengths should be understood as physical and therefore have
to be multiplied by the cosmological scale factor $a(t)$.  \\
ii. The time coordinate should be replaced by the retarded time which means
taking into account the finite velocity of light or doing the calculations on
the light cone. \\
This justifies the recent modifications to the Newtonian
calculations of different cosmological parameters related to the
structure formation which are coined with the term 'light cone
effects', see \cite{nis99}, \cite{mat97}, \cite{nak98}.
\\
The expression (\ref{gg}) may also be written in the differential form. By
taking the divergence of (\ref{gg}) and ignoring terms of the
order of $\frac{\lambda}{d_H}$, we obtain:
\begin{equation}
\nabla\cdot g=4\pi G \rho \delta,
\end{equation}
which is of the same form as the Newtonian Poisson equation except for the
modifications discussed above.   \\
The corresponding continuity equation can be obtained through the
Bianchi identities. We therefore consider the conservation of
energy-momentum tensor in the harmonic gauge. Now, using 
(\ref{T0}--\ref{T2}), the zero component of the Bianchi identities
as the conservation of the energy-momentum,
\begin{equation}
T^{\mu \nu}{}_{;\mu} = T^{\mu \nu}{}_{,\mu} + \Gamma^{\nu}{}_{\alpha\beta}
T^{\alpha\beta}+ \Gamma^{\alpha}{}_{\alpha\beta}T^{\nu\beta}=0,
\end{equation}
leads to
\begin{equation}
\label{delta}
 \dot\delta+\nabla\cdot{u} -
\frac{3}{2}h_{00,0}+h_{0}^{i}{}_{,i}=0.
\end{equation}
The third and fourth terms are small relative to the first and
second term. This can best be seen by rewriting expression
(\ref{delta}) in the Fourier space:
\begin{equation}
\dot\delta_k+\frac{ik}{a}\cdot{u_k}-\frac{3}{2}h^{(k)}_{00,0}+\frac{ik}{a}
h^{(k)}{}_{0}^{i} = 0.
\end {equation}
In the linear regime, the first term is proportional to
$\frac{\dot a}{a}\delta_k$. From expression (\ref{peebles}) it is easily
seen that the second term is of the same order of magnitude as the
first one. To estimate the third term we note first that
$h^{(k)}_{00}$ is essentially the potential energy obeying the
Newtonian relation
$$\nabla^2\phi= 4\pi G \rho\delta.$$
Taking the time derivative of the corresponding equation in
Fourier space we easily obtain:
\begin{equation}
h^{(k)}_{00,0} = 3(\frac{\dot a}{a})^2\lambda^2\dot{\delta_k}-
3(\frac{\dot a}{a})^3{\lambda}^2\delta_k
\end {equation}
The two terms on the right hand side of this equation are of the
order of $(\frac{\lambda}{d_H})^2 H \delta_k$. Therefore, the
third term in Eq.(\ref{delta}) is of the order of
$(\frac{\lambda}{d_H})^2$ times the first or the second term
and is therefore ignorable. \\
To estimate the forth term in (\ref{delta}) we consider the
${i0}$ component of the Einstein equation by taking into account
that $\lambda<d_H$:
\begin {equation}
-\frac{1}{2}h_{i0},{}^{\mu}_{\;\;\mu} = 8\pi G\rho(\frac{1}{2}h_{i0}-u_i)
\end{equation}
Taking the Fourier transformation of the above equation and using
the expression (\ref{peebles}), we obtain:
\begin {equation}
\frac{kh^{(k)}{}_{i0}}{a}\simeq (\frac{\lambda}{d_H})^2 H\delta_k,
\end{equation}
which is small relative to the first and second term and can be
ignored. Therefore (\ref{delta}) may be written as:
\begin{equation}
\label{u}
 \dot\delta+\nabla\cdot{u} = 0.
\end{equation}
We have therefore seen that by perturbing the matter dominated FRW
universe in harmonic gauge the same Newtonian hydrodynamic Eqs
.(\ref{1}) and (\ref{2}) are obtained with the two modifications
discussed above. Combining the (\ref{retard}) and
(\ref{u}) and using the fact that in the linear regime
$\delta\propto a$ we obtain as follows:
\begin {equation}
\label{27}
\nabla\cdot v(r)=\frac {H(t)f(\Omega)}{4\pi G
\rho(t)}\nabla\cdot g(r).
\end {equation}
Omitting $\nabla$ from both sides and substituting $g$ from
(\ref{retard}), we obtain for the peculiar velocity
\begin {equation}
\label{rel}
\overrightarrow{v}(r,t)=\frac{H(t) f(\Omega )}{4\pi G
b\rho(t)} \int G\rho(t-\frac{\left|r-r'\right|}{c})\delta
(r',t-\frac{\left|r-r'\right|}{c}) \frac{(\vec{r'} -
\vec{r})}{\left|\vec{r} - \vec{r}\right|^3}d^3r'.
\end {equation}
In the following sections we will rewrite the above relativistic expression
in terms of observational parameters and will compare it to the corresponding
Newtonian results.

\section{Relativistic Derivation of the Peculiar Velocity From
Density Contrast}
Assume a cluster of galaxies, $G_1,G_2,...G_n$,
at a large distance from us. Take two typical galaxies $G_1$ and
$G_2$ (Fig.1). For the purpose of calculating the relation
between density contrast and peculiar velocity we need the
gravitational action of $G_2$ on $G_1$. The situation is best
visualized in a space--time diagram. The world lines of $G_1$ and
$G_2 $ cross our past light-cone at the observation event $O$ at
$\it P$ and $\it Q$ respectively. These events correspond to our
observation of the galaxies. But the action of $G_2$ on $G_1$ is
defined through another event: crossing of the past light-cone of
${\it P}$ and the world-line of $G_2$. Call this event {\it R
}(Fig.1). Times corresponding to each of the events $\it {P, Q}$,
and ${R}$ are characterized by the corresponding subscripts. Our
observation time at the event $O$ is given by $t = t_{0}$. Now,
we have to calculate the relation between $v_{pec}$ and $\delta
(x)$ using observed data
on $\it P$ and $\it Q$.\\
Note first the following relations between time and space coordinates of
the events $\it P$ and $\it Q$, taking again $c=1$:
\begin {eqnarray}
\label{tp}
t_{\it P} = t_{0}- {\left| \overrightarrow{r_{\it P}}\right| },\\
\label{tq}
 t_{\it Q} = t_{0}- {\left| \overrightarrow{r_{\it
Q}}\right|}.
\end {eqnarray}
Let $G_1$ be the reference galaxy. The action of any other galaxy, like
$G_2$, on $G_1$ is along the past light-cone of events on the world line of
$G_1$. It is easily seen that the time $G_2$ acts on $G_1$ is given by
\begin{equation}
 t_{R} =t_{0}-\left|\overrightarrow{r_{P}}\right| -
\left|\overrightarrow{r_{R}}-\overrightarrow{r_P} \right| ,
\end{equation}
where $r_{R}$ is defined as the space coordinate of the event $R$.
Within the approximation we are interested in, one can replace $r_{R}$ by
$r_Q \simeq r_{R}$. Therefore the above relation can be written as
\begin {equation}
\label{tr}
t_{R}= t_0-\left| \overrightarrow{r_P}\right| -
\left|\overrightarrow{r_Q}-\overrightarrow{r_P}\right|.
\end {equation}
Now, Eq.(\ref{rel}) can be written in terms of space time
coordinates at $P$ and  $R$:
\begin {equation}
\label{vv}
 \overrightarrow{v}(r_P,t_P)=\frac{H(t_P) f(\Omega
)}{4\pi G b\rho(t_P)} \int d^3r_{R} G\rho(t_R)\delta
(r_{R},t_{R})\frac{(\vec{r_R} - \vec{r_P})} {\left|\vec{r_R} -
\vec{r_P}\right|^3}.
\end {equation}
Quantities depending on $R$ have to be reformulated as functions
of space time coordinates of $Q$. This is done in two steps.
First we expand $H(t_P)$ in terms of $H_{0}$ and ${\rho}(t_R)$ in
the terms of ${\rho}(t_P)$:
\begin{eqnarray}
\label{h}
H(t_P)=H_{0}(1+\frac{\left|\vec r_P\right|}{{H_0}^{-1}})\\
\label{rho}
 \rho({t_R}) = \rho({t_P})(1 + 2 \frac{\left|\vec
r_{R}-\vec r_{P}\right|} {{H_0}^{-1}}),
\end{eqnarray}
where we have used the FRW equations
\begin{eqnarray}
(\frac{\dot a}{a})^2+\frac{k}{a^2}=\frac{8\pi G}{3} \rho\\
\dot\rho+3H(\rho+P)=0
\end{eqnarray}
Substituting now the Eqs.(\ref{h}) and (\ref{rho}) into
(\ref{vv}), we obtain:
\begin {eqnarray}
\overrightarrow{v}(r_P,t_P)&=&{\frac{H_0 f(\Omega )}{4\pi b}}\int
\delta (r_R,t_R)\frac{(\vec{r_R} - \vec{r_P})}{\left|\vec{r_R} -
\vec{r_P}\right|^3}d^3r_{R} \nonumber\\
&+&\frac {H_{0}\left|\vec r_{P} \right|}{c}\frac{H_0 f(\Omega )}{4\pi b}\int
\delta (r_R,t_R)\frac{(\vec{r_R} - \vec{r_P})}{\left|\vec{r_R} -
\vec{r_P}\right|^3}d^3r_{R} \nonumber\\
&+&\frac{3H_{0}}{c}\frac{H_{0}f(\Omega)}{4\pi b}\int
\delta (r_R,t_R)\frac{(\vec{r_R} - \vec{r_P})}{\left|\vec{r_R} -
\vec{r_P}\right|^2}d^3r_{R}.
\end{eqnarray}
Note that the range of integration is taken as large as desired.
Now, terms depending on the event $R$ have to be replaced by
those depending on $Q$. Denoting the relative velocity of $G_2$
with respect to us as $\vec{V}$, we obtain
\begin{equation}
\label{rr1}
 \overrightarrow{r_R}=\overrightarrow{r_Q}-\vec{V}(t_{Q}-t_R),
\end{equation}
where
\begin{equation}
\label{v}
 \vec{V} = H_{0}\overrightarrow{r_Q}
+\overrightarrow {v_{pec}}-\overrightarrow{v_{0}}.
\end{equation}
Here $v_0$ is the peculiar velocity of us with respect to CMBR and
$v_{pec}$ is the peculiar velocity of $G_2$. Using the relations
(\ref{tq}) and (\ref{tr}) and (\ref{v}), the expression (\ref{rr1}) may be
written in the form:
\begin{equation}
\label{rr}
\vec r_R = \vec r_Q +(v_0-v_{pec}-H_0
r_Q)(\left|r_P\right|+ \left|r_P-r_Q\right|-\left|r_Q\right|).
\end {equation}
For the large scales we are considering, the term proportional to
$v_0-v_{pec}$ can be neglected relative to $H_0 r_Q$. We may also
use the following relation which is easily understood:
\begin{equation}
\frac{1}{\left |r_R-r_p\right|^3} =\frac{1}{\left
|r_Q-r_p\right|^3} \left[1+{3H_0 \left|r_Q\right|}(1-\frac{\vec
r_Q \cdot \vec r_P} {\left |r_Q-r_p\right|^2})\right].
\end {equation}
Now, let us expand the density contrast $\delta (t_{R},\vec
r_{R})$ around $(t_{Q},\vec r_{Q})$:
\begin{equation}
\label{d3}
 \delta (t_R,\vec r_{R})=\delta (t_{Q},
\overrightarrow{r_Q})-\frac {\partial\delta
(t_{Q},\overrightarrow{r_Q})}{\partial {t}}(t_{Q}-t_R)+\nabla
\delta (t_{Q},\overrightarrow{r_{Q}})\cdot
(\overrightarrow{r_{R}}-\overrightarrow{r_{Q}})+...
\end {equation}
Assuming the density contrast to be proportional to the scale
factor $a(t)$, it is seen that the time derivation of the density
contrast is given by:
\begin{equation}
\label{de}
 \frac {\partial\delta (t_{Q},\vec{r_Q})}{\partial
{t_Q}} = \frac 23\frac{\delta (t_{Q},\vec{r_Q})}{t_Q}.
\end{equation}
Substituting (\ref{de}) into (\ref{d3}), we obtain:
\begin {equation}
\delta (t_R,\vec{r_R})=\delta (t_{Q},
\vec{r_Q})-\frac 23\frac{\delta (t_Q,
\vec{r_Q})}{t_Q}(t_Q-t_R)+\nabla\delta\cdot (\vec r_R-\vec r_Q).
\end {equation}
Substituting for $t_R$ and $t_Q$ from Eqs.(\ref{tq}) and (\ref{tr})
, we obtain finally:
\begin{eqnarray}
\delta (t_{R},\vec{r_R})&=&\delta (t_Q,
\vec{r_Q})-\frac{2\delta (t_Q,
\vec{r_Q})}{3{H_0}^{-1}}
(\left| r_P\right| +\left| r_P-r_Q\right|-\left| r_Q\right| )
-{\nabla\delta(t_Q,r_Q)\cdot v}\nonumber\\
&\times&(\left| r_P\right| +\left|
r_P-r_Q\right|-\left| r_Q\right| )
-\frac{\vec r_Q\cdot\nabla\delta(t_Q,r_Q)}{H^{-1}}
(\left| r_P\right| +\left| r_P-r_{Q}\right|-\left| r_{Q}\right|).
\end{eqnarray}
We have still to write the measure of the integral in
(\ref{vv}), $d^3 r_{R}$, in terms of the observed volume
$d^3r_{Q}$:
\begin {equation}
d^3r_{R}=\left\| \frac{\partial r_{R}}{\partial r_{Q}}
\right\| d^3r_{Q}.
\end {equation}
From Eq.(\ref{rr}), ignoring higher order terms, the Jacobian is
obtained to be:
\begin {eqnarray}
\left\| \frac{\partial r_R}{\partial r_Q}\right\|&=&1-{
\nabla \cdot v}(\left| r_P\right| +\left| r_P-r_Q\right|
-\left| r_Q\right| ) + {\vec{v}
\cdot\vec{k}}\nonumber\\
&-&{3H_{0}}(\left| r_P\right|+\left| r_P-r_Q \right|
-\left| r_Q \right|) + H_{0}{\vec r_Q\cdot\vec k},
\end {eqnarray}
where
\begin {equation}
k^i = \frac{r_Q^i}{\left| r_Q\right| } - \frac{r_{Q}^i-x_P^i}{\left| r_Q-x_P
\right| }.
\end {equation}
It can easily be seen that, within the approximation we are
considering, it is allowed to replace in the above relation the
divergence of the velocity by the density contrast from
expression (\ref{peebles}) to obtain the final result:
\begin {eqnarray}
d^3r_R&=&(1+\frac{f(\Omega )}{bH_0^{-1}}\delta
(r_Q,t_{Q})(\left| r_P\right| +\left| r_P-r_Q\right| -\left| r_Q\right| )
-{\overrightarrow {v}\cdot\overrightarrow {k}}\nonumber\\
&-&{3H_{0}}(\left| r_P\right|+\left| r_P-r_Q \right|
-\left| r_Q \right|)-H_{0}{\vec r_Q\cdot\vec k})d^{3}r_Q.
\end {eqnarray}
Replacing now all terms corresponding to the point $R$ with those defined at
the point $Q$, we obtain finally for the peculiar velocity
\begin{equation}
\label{vvv}
 \vec{v}(r_P) = \frac{H_{0}f(\Omega )}{4\pi b}\int
\frac{\delta (t_Q,r_Q)}{\left| r_P-r_Q\right| ^3} (\vec r_Q-\vec
r_P)d^3r_Q + \vec{G(r_P)},
\end{equation}
where
\begin{eqnarray}
\label{g}
 \vec G(r_P) &=&\vec{F_{1}}(r_P)+\vec{F_{2}}(r_P)+\vec{F_{3}}(r_P)
+\vec{F_{4}}(r_P)+\vec{F_{5}}(r_P)\nonumber\\
&+&\vec{F_{6}}(r_P)+\vec{F_{7}}(r_P)+\vec{F_{8}}(r_P)
+\vec{F_{9}}(r_P),
\end{eqnarray}
and
\begin{eqnarray}
\label{f1}
 \vec{F_1}(r_P)&=& \frac{ H_0^2 f(\Omega )}{4\pi b}\int
\delta (r_{Q},t_{Q}) \frac{(\vec r_Q - \vec r_P)}{\left|\vec r_Q
-\vec r_P\right|^3} (\left| r_P\right| +\left| r_Q-r_P\right|-
\left| r_Q\right|)
(\frac{\left |r_Q \right|^2}{\left |r_Q-r_P \right|^2}\nonumber\\
&-&\frac{\vec r_P \cdot \vec r_Q}{\left |r_Q-r_P \right|^2})d^3r_{Q},\\
\vec{F_2}(r_P)&=&\left|\vec r_{P} \right|
\frac{ H_0^2 f(\Omega )}{4\pi b}\int \delta (r_{Q},t_{Q})
\frac{(\vec r_Q - \vec r_P)}{\left|\vec r_Q -\vec r_P\right|^3}d^3r_{Q},\\
\vec{F_3}(r_P)&=&\frac{2 H_0^2 f(\Omega )}{4\pi b}\int
\delta (r_Q,t_Q)\frac{(\vec {r_Q} - \vec {r_P})}{\left|\vec {r_Q} -
\vec {r_P} \right|^2} d^3r_{Q},\\
\vec{F_4}(r_P)&=&-\frac{11H_0^2f(\Omega )}{3\pi b}\int
\frac{\delta (t_Q,r_Q)}{\left| r_P-r_Q\right| ^3}
(\vec{r_Q}-\vec{x_P})(\left|
r_P\right| +\left| r_Q-r_P\right| -\left| r_Q\right| )d^3r_Q,\\
\vec{F_5}(r_P)&=&-\frac{H_0^2f(\Omega )}{4\pi b}\int
\frac{\vec{r_Q}\cdot \nabla \delta (t_Q,r_Q)}
{\left| r_P-r_Q\right| ^3}(\vec{r_Q}-
\vec{r_P})(\left| r_P\right| +\left| r_Q-r_P\right| -\left|
r_Q\right| )d^3r_Q,\\
\vec{F_6}(r_P)&=&-\frac{H_0^2f(\Omega )}{4\pi b}\int
\frac{\delta (t_{Q},r_Q)}{
\left| r_P-r_Q\right| ^3}(\left| r_P\right| +\left| r_Q-r_P\right| -
\left|r_Q\right| )\vec {r_Q}d^3r_Q,\\
\vec{F_7}(r_P)&=&-\frac{H_0f(\Omega )}{4\pi b}\int
\frac{\delta (t_Q,r_Q)}{
\left| r_P-r_Q\right| ^3}(\left| r_P\right| +\left| r_Q-r_P\right| -\left|
r_Q\right| )\vec{v(r_Q)}d^3r_Q,\\
\vec{F_8}(r_P)& = &+\frac{H_{0}f(\Omega )}{4\pi b}\int \frac{\vec{v(r_Q)}
\cdot\vec{k}}{\left| r_P-r_Q\right| ^3}(\vec r_Q-\vec r_P)
\delta(t_Q,r_Q)d^3r_Q,\\
\label{f9}
 \vec{F_{9}}(r_P)&=&-\frac{H_0f(\Omega )}{4\pi b}\int
\frac{\vec{v(r_Q)}\cdot\nabla \delta (t_Q,r_Q)}{\left|
r_P-r_Q\right| ^3}(\vec{r_Q} -\vec{r_P})(\left| r_P\right|
+\left| r_Q-r_P\right| -\left| r_Q\right| )d^3r_Q.
\end{eqnarray}
To simplify the results and to have an estimation of the
corrections to the Newtonian expressions, we take a closer look
at different terms in (\ref{g}). Let us first distinguish
between $F_1 - F_4$, and $F_6$ on one side and $F_7 - F_9$ on the
other side. The latter terms are typically of the order of formers
times the fraction of the peculiar velocity divided by the Hubble
velocity, being of the order of $10^{-2}$. Therefore, the latter
terms may be neglected. The term $F_5$ contains the factor $\vec
r_Q \cdot \nabla \delta$ which is vanishing in general. To see
this, consider a cluster around $r_Q$. Because of the symmetric
distribution of matter in the cluster the gradient term accept
positive and negative values while $r_Q$ is almost constant.
Therefore, one may assume that the integral in $F_5$ is vanishing.
The same is true about the contribution of the term $\frac{\vec
r_Q\cdot \vec r_P}{\left|\vec r_P - \vec r_Q \right|^2}$ in
$F_1$. This is due to the large extent of our integration domain
represented by $r_Q$. Now, for Stromlo--APM-- and Las
Campanas--red-shift survey we may assume the distance to the
cluster to be of the same order as the extent of the cluster,
$L$, so that we have  $\left |\vec {r}_P - \vec{r}_Q \right|
\simeq \left| \vec{r}_P\right| \simeq \left| \vec{r}_Q\right|
\simeq L \simeq 100 Mpc$. A closer look at the remaining terms
shows that $F_{1, 3, 4, 6}$ are of the order of or bigger than
$F_2$. We may therefore write
\begin{equation}
\vec G \simeq \frac{4LH_0}{3}\vec W,
\end{equation}
where
\begin{equation}
\vec W = \frac{H_{0}f(\Omega )}{4\pi b}\int
\frac{\delta (t_Q,r_Q)}{\left| r_P-r_Q\right| ^3}
(\vec r_Q-\vec r_P)d^3r_Q.
\end{equation}
Substituting now $\vec {G}$ in expression (\ref{vvv}) we obtain finally:
\begin{equation}
\vec v \simeq \vec W(1 + \frac{4L}{3c H_{0}}),
\end{equation}
where $L$ is the size of structure and $c$ is explicitly inserted
again. In the limit $c \rightarrow \infty$ or $L\rightarrow 0$ 
we obtain the
Newtonian value for the peculiar velocity, $v_N$, which is just
the first term on the right hand side of (\ref{vvv}).
Therefore, the relative relativistic correction to the peculiar
velocity up to the first order of $\frac{1}{c}$ is given by
\begin{equation}
\label{vrelat}
 \frac{v_{rel}-v_N}{v_N} = \frac{4}{3}\frac{LH_{0}}{c},
\end{equation}
where we have taken the absolute values for the velocities and
added the subscripts $_{rel}$ and $_N$ to emphasize the
relativistic corrections. Taking the divergence of the
(\ref{vvv}), we obtain a relation between the density contrast
and the divergence of the peculiar velocity:
\begin{equation}
\label{divv}
 \frac{b}{H_{0}f(\Omega)}\nabla\cdot \vec{v}(r_P) =
\delta(r_P) - \frac{b}{c H_{0}f(\Omega)}\nabla\cdot \vec{G}(r_P).
\end {equation}
Here again, in the limit $c \rightarrow \infty$ the second term
on the right hand side vanishes and we get the familiar Newtonian
expression for the density contrast, which will be denoted by
$\delta_N (r_P)$. Now, to distinguish the density contrast
appearing on the right hand side of (\ref{divv}) from the
Newtonian value, we call it $\delta_{rel}$. The expression (\ref{divv}) is
now written as:
\begin{equation}
\label{drel}
 \delta_{rel}(r_P) = \delta_N(r_P) +
\frac{b}{H_{0}f(\Omega)c} \nabla\cdot \vec{G}(r_P).
\end {equation}
In the following sections we will calculate the density and
velocity power spectra for both Newtonian and relativistic cases.
\section{Power Spectrum: Newtonian Versus Relativistic Case}
Take the correlation function of density contrast as
\begin {equation}
\label{corr}
 \xi(r)=\int\delta(r+r')\delta(r')\frac{d^3r'}{V},
\end{equation}
defined in a cosmological volume $V$.  For simplicity we will omit hereafter
the subscript $_P$ in $r_P$. The power spectrum is given by the
Fourier transform of the correlation function:
\begin {equation}
\xi(r)=\int P(k)e^{ikr}d^3k.
\end{equation}
Using relation (\ref{drel}) between the Newtonian and relativistic
density contrast, substituting it in Eq.(\ref{corr}) and ignoring
higher order terms in $\delta$, we obtain the following relation
between the Newtonian-- and relativistic--correlation functions:
\begin {equation}
\label{corrn}
 \xi_N(r)=\xi_{rel}(r)-\frac{b}{c
H_0f(\Omega)}\int\delta(r'+r)\nabla \cdot \vec{G}(r')d^3r'-
\frac{b}{c H_0f(\Omega)}\int\delta(r')\nabla\cdot
\vec{G}(r'+r)d^3r'.
\end{equation}
The divergence term on the right hand side is calculated in the
Appendix A and is given by
\begin {equation}
\label{nab}
 \nabla\cdot \vec{G}(r) =
\frac{-4H_0^2f(\Omega)}{3b}\delta(r,t) \left| r \right|
\end {equation}
Using this relation and taking the Fourier transform of
expression (\ref{corrn}), the following relation between the Newtonian--
and relativistic--power spectrum is easily obtained:
\begin {equation}
P_N(k)=P(k)_{rel}(1+\frac{16 \pi H_0}{3ck}).
\end {equation}
In the limit $c \rightarrow \infty$, the Newtonian value is regained.
\section{$\beta$--Value: Newtonian Versus Relativistic }
Let us now calculate the relativistic velocity power spectrum.
The peculiar velocity is now expressed in terms of its Fourier
components:
\begin {equation}
\label{73}
 v(r)=\frac{V}{(2\pi)^{3/2}}\int v_k e^{ikr}d^3k,
\end {equation}
where $V$ is the space volume. Its power spectrum is
defined as $P_V(k)=<\left|v_{kx}\right|^2 + \left|v_{ky}\right|^2
+ \left|v_{kz} \right|^2>$. If $v(r)$ is an isotropic Gaussian
field, then the different Fourier components are uncorrelated and
the power spectrum provides a complete statistical description of
the field. The Velocity spectrum is given by
\begin {equation}
\label{pows}
V^2(k)=\frac {1}{2 \pi^2}k^3P_V(k).
\end {equation}
As we have shown in the appendix B, the Fourier transform of
(\ref{vrelat}) leads then to the following relation between the
Newtonian-- and relativistic-- velocity spectrum:
\begin {equation}
\label{mm}
V^2_{rel}(k)=V^2_{N}(k)(1+\frac{16 \pi H_0}{3ck}).
\end {equation}
The Newtonian spectrum is independently obtained by taking the
Fourier transform of (\ref{peebles}):
\begin {equation}
\label{vn}
 V^2_{N}(k) = \frac{1}{2 \pi^2}\beta^2 H_0^2kP(k).
\end{equation}
where $\beta = \frac{f(\Omega)}{b}$. These relations are obtained
in terms of real space expressions in contrast to power spectrum
data which are obtained in the red-shift space \cite{kai87}. It is
used to change the factor $\beta^2$ in front of the term on the
right hand side of (\ref{vn}) to $F(\beta)$ to account for the
transformation between red-shift- and real-space. Therefore we may
write the final equation in terms of red-shift data in the form
\begin {equation}
V^2_{N}(k) = \frac{1}{2 \pi^2} F_N^2(\beta) H_0^2kP(k),
\end{equation}
where
\begin{equation}
F_N^2(\beta) = \frac{\beta^2}{1+2\beta/3+\beta^2/5}.
\end{equation}
Now, using expression (\ref{mm}) the relativistic velocity spectrum will
be given in the form
\begin{equation}
V^2_{rel}(k) = \frac{1}{2 \pi^2}F_{rel}^2(\beta) H_0^2kP(k),
\end {equation}
where
\begin{equation}
F_{rel}^2(\beta) = F_N ^2(\beta) (1 + \frac{16\pi H_0}{3ck}).
\end {equation}

\section{Discussion}
Any observation in cosmology is carried along the light cone and
not on a space-like slice defined by a definite 
observer time. However the difference between the two procedures
are usually ignorable, which leads to the usual ignorance of the
light cone effects and the finite signal velocity. At
cosmological distances this difference can leads
however to observable effects. \\
Here we have calculated the peculiar velocity--density contrast relation
in a cluster of galaxy taking into account the finite signal velocity. The
result differs from the familiar Newtonian relation through a lengthly term
which can however be simplified for a typical cluster. The estimation based
on this simplification depends on the extend of the cluster; the larger the
extent the bigger the difference.
\subsection*{ Appendix A}
Let us calculate the term $\nabla\cdot \vec{G}(r)$ appearing in
(\ref{drel}) and (\ref{corrn}). Using eqs.(\ref{f1}-\ref{f9}) we
may write:
\begin{eqnarray}
\nabla \cdot \vec{G}(r_P) &=& \frac{3 H_0^2 f(\Omega)}{4\pi b}\int\frac{\vec r_P
\cdot \vec r_Q}{\left| r_P \right| \left| r_P-r_Q \right|^3}
\delta(r_Q)d^3r_Q - \frac{3 H_0^2 f(\Omega)}{4\pi b}\int\frac
{\left| r_P \right|}{\left| r_P-r_Q \right|^3}\delta(r_Q)d^3r_Q\nonumber\\
&-& \frac{4 H_0^2f(\Omega)}{3b}\left| r_P \right|\delta(r_P),
\end{eqnarray}
The symmetric distribution of matter in large scales leads to
positive and negative values for $\vec r_P\cdot \vec r_Q$ in the
first integral which makes it vanishing. The second integral on
the right hand side may be written as $\frac{3H_0^2 f(\Omega)}{4
\pi b}\left | r_P \right | \int\frac{\delta(r_Q)} {\left|
r_P-r_Q\right|^3}d^3r'\simeq\frac{3H_0^2 f(\Omega)}{4 \pi b}
<\delta>L$. The average value of density contrast, $<\delta>$,
taking the domain of integration sufficiently large, does vanish
too. We then obtain finally:
\begin {equation}
\nabla \cdot \vec{G}(r_P) = - \frac{4 H_0^2f(\Omega)}{3b}\left| r_P \right|
\delta(r_P).
\end {equation}
\subsection*{Appendix B }
In order to derive the Newtonian power spectrum we substitute
expression (\ref{nab}) into (\ref{corrn}):
\begin {equation}
\xi_{N}(r) = \xi_{Rel}(r)+\frac{8\pi H_0}{3c}\int \delta(r'+r)\delta(r')
\left| r' \right| d^3r'.
\end {equation}
Taking the Fourier transformation of it we obtain
\begin {eqnarray}
\int P_{N}(k) e^{ik.r}d^3k &=&  \int P_{Rel}(k) e^{ik.r}d^3k\nonumber\\
&+& \frac{1}{(2\pi)^3}\frac{8 H_0}{3c}
\int \delta_k \delta_k'e^{ik.(r+r')}e^{ik'.r'}r'd^3r'd^3kd^3k'.
\end {eqnarray}
Putting for simplicity $r'= \frac{2 \pi}{k}$ we may write
$\int e^{ir'.(k+k')}d^3r'=(2 \pi)^3\delta^3(k+k')$, which leads to
\begin {equation}
P_{N}(k) = P_{Rel}(k)(1+\frac{16 \pi H_0}{3kc}).
\end {equation}
Calculation of the velocity power spectrum is similar to above
procedure. One start with (\ref{vrelat}), by substituting it
into (\ref{73}) and taking Fourier transformation of it and using
Eq.(\ref{pows}) , it is easily seen that:
\begin {equation}
V^2_{rel}(k)=V^2_N(k)(1+\frac{16 \pi H_0}{3ck}).
\end {equation}

\section{Figure Captions}
Fig. 1. Our past light cone crosses the world-lines of galaxies
$G_1$ and $G_2$ at $P$ and $Q$ respectively. The event $R$, on the
intersection of the world line of $G_2$ and the past light cone
of $P$, gives the moment at which $G_2$ acts on $G_1$ being effective
at $P$.\\
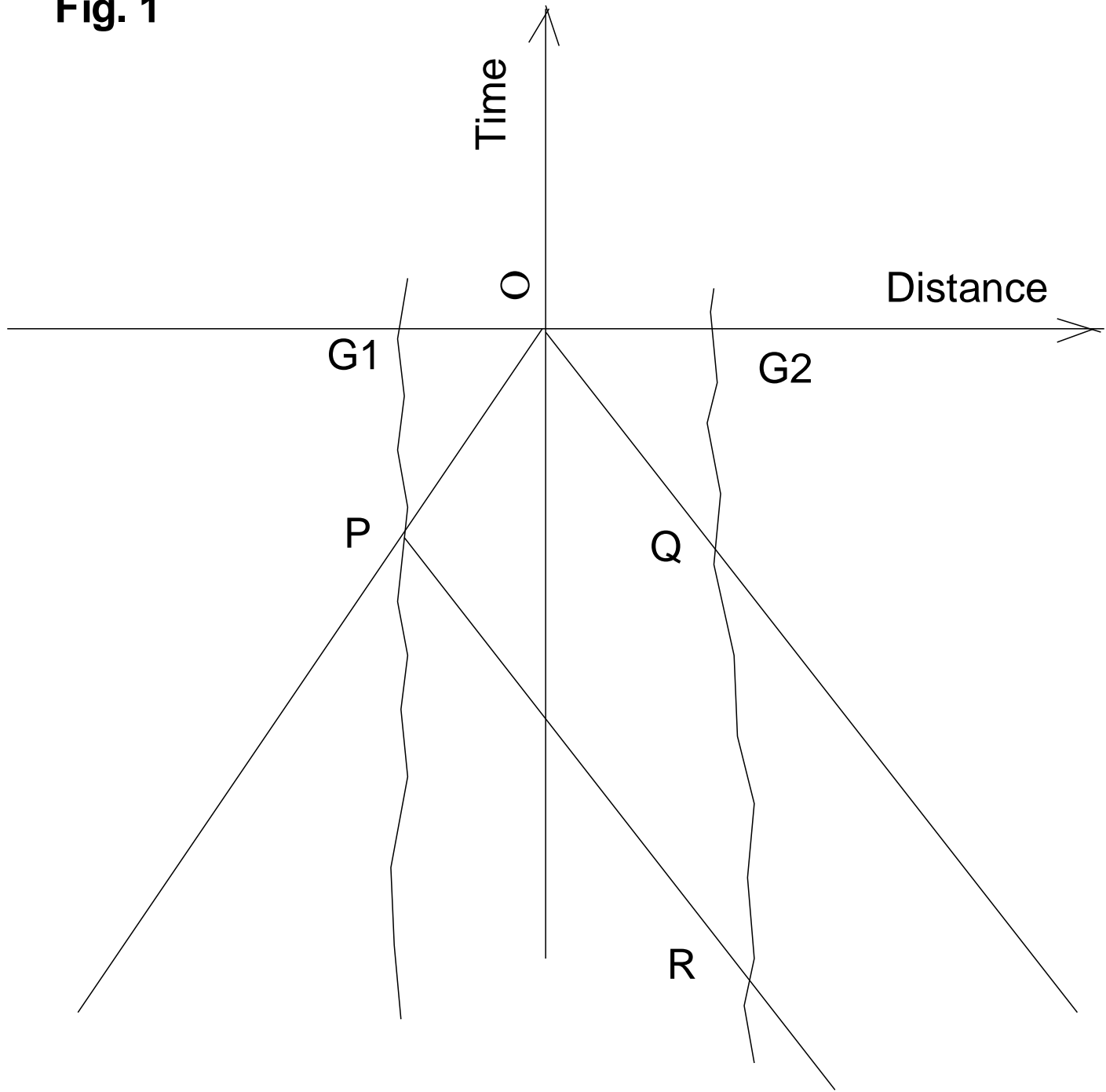
\end{document}